# El impacto del marketing digital en empresas fabricantes de embutidos de los Altos de Jalisco

*The Impact of Digital Marketing on Sausage Manufacturing Companies in the Altos of Jalisco*

*O impacto do marketing digital nas empresas fabricantes de salsichas dos Altos de Jalisco*

**Guillermo José Navarro del Toro**
Universidad de Guadalajara, México
navarromemo@hotmail.com
guillermo.ndeltoro@academicos.udg.mx
https://orcid.org/0000-0002-4316-879X

**Resumen**

Uno de los objetivos de cualquier empresa, además de elaborar productos de alta calidad y aceptación comunitaria, es elevar las ventas de manera significativa. Desgraciadamente, existen regiones donde aún no se utilizan las nuevas tecnologías de mercadeo que hacen posible llegar a una mayor cantidad de posibles consumidores, no solo a nivel regional, sino también estatal y nacional. La presente investigación, que incluyó métodos cualitativos y cuantitativos, así como entrevistas aplicadas a propietarios, empleados y clientes de tres empresas de embutidos, busca sondear el impacto del *marketing* digital en la región de los Altos de Jalisco, México. Así, además de indagar sobre el grado de conocimiento que tienen respecto de las tecnologías de la información y comunicación (TIC) para ampliar sus mercados hacia zonas de mayor densidad poblacional, se busca conocer la opinión sobre sus productos elaborados, su calidad y aceptación. No hay que olvidar que las empresas están transitando a un mundo cada vez más conectado, lo que posibilita que los empresarios pueden hacer llegar sus productos a una mayor cantidad de consumidores, mediante Internet y dispositivos inteligentes, tales como celulares, tabletas y computadoras, y así asegurar la supervivencia de la empresa y una permanencia más prolongada en el mercado.

**Palabras clave:** empresa, estrategias de *marketing*, herramientas digitales, Internet, medios digitales.

## Abstract

One of the goals of any business, in addition to producing high-quality, community-accepted products, is to significantly increase sales. Unfortunately, there are regions where new marketing technologies that make it possible to reach a larger number of potential consumers, not only at the regional level, but also at the state and national level, are not yet used. This research, which included qualitative and quantitative methods, as well as interviews applied to owners, employees and clients of three sausage companies, seeks to measure the impact of digital marketing in the Altos of Jalisco, Mexico. Thus, in addition to inquiring about the degree of knowledge they have regarding information and communication technologies (ICT) to expand their markets to areas with higher population density, another goal is to know the opinion about their manufactured products, their quality and acceptance. It should not be forgotten that companies are moving to an increasingly connected world, which enables entrepreneurs to get their products to a greater number of consumers through the Internet and smart devices, such as cell phones, tablets and computers; and thus ensure the survival of the company and a longer stay in the market.

## Resumo

Um dos objetivos de qualquer empresa, além de produzir produtos de alta qualidade aceitos pela comunidade, é aumentar significativamente as vendas. Infelizmente, há regiões onde ainda não são utilizadas novas tecnologias de marketing que permitem atingir um maior número de consumidores potenciais, não só no âmbito regional, mas também estadual e nacional. Esta pesquisa, que incluiu métodos qualitativos e quantitativos, além de entrevistas aplicadas a proprietários, funcionários e clientes de três empresas de embutidos, busca sondar o impacto do marketing digital na região de Los Altos de Jalisco, no México. Assim, além de indagar sobre o grau de conhecimento que possuem em relação às tecnologias de informação e comunicação (TIC) para expandir seus mercados a áreas com maior densidade populacional, buscamos saber a opinião sobre seus produtos manufaturados, sua qualidade e aceitação. Não se deve esquecer que as empresas caminham para um mundo cada vez mais conectado, que permite aos empreendedores levar seus produtos a um maior número de consumidores, por meio da internet e de dispositivos inteligentes, como celulares, tablets e computadores, e garantindo assim a sobrevivência da empresa e uma maior permanência no mercado.

# Introducción

El objetivo principal de la presente investigación es conocer, de manera aproximada, el grado de impacto que tienen las nuevas tecnologías en la región de los Altos de Jalisco, específicamente el *marketing* digital, el cual implica el uso de tabletas, teléfonos inteligentes, computadoras y enlaces de comunicación (de forma alámbrica e inalámbrica). Debido a que, a través del Internet, es posible contribuir a que los productos de las empresas de dicha zona puedan llegar a un mercado mayor. Si se toma en cuenta la calidad de muchos de ellos, puedan encontrar un nicho en zonas como el Área Metropolitana de Guadalajara (AMG). Además, mucha gente oriunda de los Altos acude frecuentemente ahí con motivos diversos y a su regreso trae consigo una gran cantidad de productos regionales. Así pues, esta investigación se limitó a las empresas alteñas dedicadas a la fabricación de embutidos, para descubrir los motivos por los cuales siguen sin expandirse y llegar a ese público que, siendo nativo del área, solamente pueden consumir los productos que allí se elaboran, única y exclusivamente, cuando visitan la región.

Más específicamente se tomaron en cuenta solo tres empresas de este giro, las más representativas de los Altos (en lo referente a su aceptación, cobertura y tamaño), a saber, Embutidos Vera, El Chaparral y Embutidos Navarro, para poder estandarizar un criterio que permita conocer los motivos por los cuales sus líneas de distribución continúan restringidas a un área relativamente pequeña. Este estudio está enfocado en conocer el grado de participación que tienen los nuevos métodos de *marketing* en esas empresas y su penetración en los mercados de la región, así como el grado de relación que tienen los medios de publicidad tradicional con esa mínima expansión. Respecto a estos últimos, uno de los métodos publicitarios más tradicionales es el denominado *difusión de boca en boca*, el cual es definido como "un conjunto de actividades de la empresa que buscan dar al consumidor motivos para hablar de sus productos o servicios, y proporcionarle las herramientas adecuadas para que esas conversaciones se produzcan muchas veces" (MarketingDirecto.com, s. f.).

En cuanto a las empresas visitadas como parte de este estudio, las tres datan de la última década del siglo pasado, dos de ellas son locales y la otra proviene de la población de Arandas. Todas coinciden en ser fabricantes, distribuidoras y vendedoras del mismo producto: embutidos (longaniza). De manera paulatina, todas ellas, fueron introduciendo en el mercado sus productos y teniendo gran aceptación por la calidad de las materias primas y por los métodos de fabricación empleados.

## Problemática

Uno de los principales problemas que enfrenta la mayoría de las empresas fabricantes de embutidos de la región de los Altos está directamente relacionado con los métodos de publicidad que emplean. Básicamente se reducen a la entrega de propaganda impresa con las ofertas que tienen y a la publicidad de boca en boca, en donde los clientes, sobre todo aquellos que los conocen como productos de calidad desde el inicio, son quienes se encargan de recomendar con base en su experiencia los productos de la empresa.

Para incrementar las ventas de esos productos, así como expandir sus mercados y obtener nichos de aceptación en zonas de alta densidad poblacional, como el AMG, se requiere implementar el *marketing* digital. Para ello, se tiene que involucrar a gente que maneje técnicas mercadológicas y herramientas de comunicación tecnológica; así se podrá proyectar tanto crecimiento como aceptación fuera de su lugar de origen.

Cabe resaltar que no solo dueños y empleados deben de estar capacitados en el manejo de las nuevas tecnologías, y sobre todo en la forma en que podrán incrementar las ventas, sino, más importante aún, es acercarse a los clientes para que se interesen en el nuevo sistema de mercadeo, para que lo acepten al ver sus bondades y ventajas sobre los tradicionales. Con esto, al tener un mayor alcance, se estará más cerca de una permanencia prolongada y un crecimiento sostenible.

## Marco de referencia
### Impacto de internet

En la última década de 1900 inició el uso de Internet en algunos países como medio de comunicación en contextos universitarios hasta expandirse e impactar las vidas de casi todo el mundo, ya que actualmente es una herramienta consolidada de amplia aceptación y uso en muchas de las sociedades modernas. Ha cambiado muchos hábitos y comportamientos, a tal grado que sería imposible concebir una comunicación efectiva y eficaz sin ella. En la mayoría de los países, es una forma habitual de tratar la información por parte de un sinnúmero de usuarios. Asimismo, se transformó en la herramienta por antonomasia para captar información, buscar contenidos escritos, audios, visuales, comprar, generar relaciones, entretenimiento o simple trabajo. Toda una revolución en la comunicación, ha desplazado a muchos de los medios tradicionales como la radio y la televisión para mantener informada a la sociedad, así como para presentar contenidos en formatos que en otro tiempo no se podrían siquiera pensar, de forma inmediata, sin esperar a que alguien informe o muestre lo que pasa en algún lado del mundo.

La revolución digital cambió sustancialmente la forma de realizar las actividades de *marketing* (mercadeo) en empresas e industrias. De hecho, el *marketing* actualmente es una filosofía. Se trata de una forma y actitud de concebir una relación de intercambio de las industrias con el entorno, enfocándose en el consumidor. La relación de intercambio determina la razón de ser y estar de las empresas en el mercado; ninguna empresa puede mantenerse en el mercado sin establecer un lazo estrecho con el consumidor, sin satisfacer sus necesidades de la forma óptima.

A diferencia del mercadeo tradicional, que tiene un horario de atención al cliente, el mundo digital está conectado 24 horas al día 7 días a la semana, sin importar el lugar en donde se localicen empresa y cliente. En tales términos, se puede incluso considerar que el *marketing* digital ha tenido un crecimiento descontrolado. Como sea que fuere, paulatinamente se van sumando estrategias empresariales para dar solución a los nuevos retos (Famet Andalucía, 2016).

# Marketing digital

El *marketing* es un concepto que engloba el conjunto de actividades destinadas a satisfacer deseos y necesidades de individuos, empresas y organizaciones a cambio de alguna remuneración (costo), por lo que se convirtió en una herramienta indispensable para incrementar las posibilidades de éxito en los mercados, ya que en él se puede incluir desde el trueque (sistema en donde se intercambian mercancías por otras) hasta el novedoso *marketing* digital (Martínez, Martínez y Parra, 2015).

El *marketing* digital se define como una aplicación (programa de computadora) que combina tecnologías digitales para contribuir y hacer más fáciles las actividades de mercadeo; su empleo, de manera general, hace rentable la adquisición de bienes y servicios, ya que se incrementa continuamente la captación de clientes. Con ello, la tecnología digital y el desarrollo del enfoque planificado adquieren un reconocimiento estratégico y permiten a las empresas mejorar su conocimiento del cliente, así como sus gustos, preferencias y productos (presentación, calidad, cantidad) con el fin de satisfacer sus necesidades.

Surgió con las primeras páginas web implementadas para promocionar productos o servicios. Con el tiempo, se incrementó el número de herramientas desarrolladas como parte de las nuevas tecnologías, y con ello se facilitó el contar con desarrollos más sofisticados y eficientes, tal como son las aplicaciones (*apps*), así como el proceso de gestión y análisis de datos recolectados del cliente. El *marketing* digital ahora es el encargado de esas dimensiones, herramienta indispensable en toda empresa (Cangas y Guzmán, 2010).

Es, pues, un recurso de alto impacto que interactúa con el consumidor directamente, hace el proceso dinámico, permite captar más información a través del Internet para generar mayor audiencia a bajo costo, reduce tiempo y costos de ventas, incrementa las ventas en línea mediante canales electrónicos de publicidad que tienen gran rapidez para mostrar lo que se puede vender. El *marketing* digital se ha posicionado como el rumbo a seguir en el desarrollo de estrategias empresariales, y, al abreviar el tiempo de respuesta de las necesidades del cliente, ha simplificado los procesos de *marketing* entre consumidor/empresa (Vargas, 2017).

Desafortunadamente, México tiene una inversión muy pobre (entre 20 y 24 dólares por usuario) en torno al *marketing* digital. Está posicionado muy por debajo de países como Estados Unidos y Reino Unido, cuya inversión oscila entre 165 y 185 dólares por usuario. El problema en México reside en la falta de cultura de innovación, por lo que es urgente que las empresas inviertan más en el desarrollo de herramientas que permitan generar estrategias para que entren en competencia con sus similares regionales, nacionales e internacionales (Kutchera, García y Fernández, 2014).

## Herramientas del *marketing* digital

El *marketing* digital se apoya en instrumentos que requieren distintos grados de conocimiento y especialización. Los más sencillos son los siguientes:

- Herramienta de automatización: son herramientas que desarrollan funcionalidades de *landing pages*,[1] *e-mail marketing* y flujos de automatización de *e-mail*. Todo esto provee y aumenta la gestión de *leads*[2] y la madurez de estos en un embudo de ventas.

- *E-mail*: es uno de los canales fundamentales de interacción con los clientes, posterior al primer contacto del consumidor con una página o empresa. Es gracias al *e-mail marketing* que se hace posible comunicarse con él y brindarle más contenido hasta que esté preparado para entrar en contacto con el equipo de ventas.

- Plataforma de contenido: estas herramientas pueden mejorar mucho el proceso porque permiten desde una gestión del blog hasta la creación de demandas para la producción de *posts*, lo que se traslada en un mayor ahorro de tiempo y mayor eficiencia.

- Herramienta de *analytics*: el cálculo del resultado de las acciones en línea es básico para obtener los mejores datos de un negocio. Se hace mediante la plataforma Analytics de Google, por ejemplo, su gran ventaja es la evaluación e interpretación del interés de los visitantes en el sitio. Aquí se mide el rendimiento de la inversión (ROI, por sus siglas en inglés). De las acciones realizadas, se detectan las estrategias que den mejores resultados para atraer e interesar al público.

- Monitorización de redes sociales: ayuda a mejorar las acciones en las redes, estimulando el crecimiento de la productividad a través de *posts*, y permite seguir las menciones de la marca, evaluar el interés y reacciones a los contenidos y la base de seguidores (Lipinski, 29 de mayo de 2020).

---

[1] Estas páginas tienen como finalidad que el usuario realice una acción: suscribirse a un boletín, compartir contenido en redes sociales, comprar un producto o suscribirse a un servicio (Díaz, 16 de noviembre de 2018).
[2] Un *lead* es un usuario que ha entregado sus datos a una empresa y que, como consecuencia, pasa a ser un registro de su base de datos con el que la organización puede interactuar. Para ello también es necesario que esta persona haya aceptado la política de privacidad de la compañía (Bel, 28 de abril de 2020).

## *Marketing* digital actual

"En esta época, los mercadólogos necesitan responder con mayor velocidad a la retroalimentación de sus usuarios o consumidores, porque los públicos están conectados 24 horas al día" (Striedinger, 2018, p. 7). Inclusive, hasta poco antes de dormir, la gente le dedica una última mirada al celular. La globalización ha traído nuevas y cada vez más eficaces herramientas, por lo que sería un gran error por parte de las empresas prescindir de los medios digitales para promocionar sus productos o servicios. Los medios informáticos trajeron grandes ventajas a las grandes empresas; ahora, gracias a estos, son todavía más competitivas a mediano y largo plazo, pueden vislumbrar un futuro más seguro y rentable. Desgraciadamente, contrario a esta tendencia, las pequeñas empresas no han obtenido gran provecho de los medios digitales, básicamente por su carencia de información, ignoran la importancia y valor agregado que traen consigo (Striedinger, 2018).

Las empresas pueden implementar estrategias de *marketing* digital para alcanzar sus objetivos y establecer una relación directa con sus clientes. A continuación, citamos algunas de ellas:

- Sitio web: se puede acceder a uno de estos a través del celular, tableta o computadora. De preferencia con un diseño llamativo, se busca que los clientes potenciales indaguen y hagan transacciones de manera frecuente. Los tipos de sitios web, corporativo y comercial, muestran información de la empresa, e inducen al cliente a comprar y realizar transacciones ahí mismo.

- Blog: de acuerdo con Arenas (2012), "es un sitio web que se actualiza periódicamente y ofrece información de uno o varios autores sobre temas de interés" (p. 22). Es un sitio donde se publican artículos relacionados con el giro de la empresa.

- Posicionamiento en buscadores: se trata de la modificación de datos en una página web con miras a alcanzar un mejor posicionamiento entre los internautas (OKhosting, 2018). Tiene como fin el colocar el sitio web en los navegadores y aparecer en los resultados de Google, Yahoo! y el resto de los buscadores.

- Redes sociales: según Celaya (2008), son lugares en Internet donde se publica y comparte todo tipo de información, personal y profesional, tanto con personas conocidas como desconocidas. Por su uso, tienen mayor alcance en el mundo, se utilizan como medios de comunicación, para emitir y recibir mensajes, ya sean videos, audios o textos, y han tenido un gran impacto en la comunicación e industria en general.

El mismo Celaya (2008) establece tres clasificaciones de redes sociales:

*1)* Profesionales (LinkedIn, Viadeo).

*2)* Generalistas (Facebook, Twitter, Instagram).

*3)* Especializadas (Ediciona, CinemaVIP, eBuga).

- Influenciadores o *social influence marketing*: esta práctica se define como "la capacidad de expandir y multiplicar un mensaje, acción o comportamiento a través de una persona con credibilidad y empatía, capaz de convencer a un grupo de personas de forma expansiva, progresiva y permanente" (Anzures, 2016, p. 137).

## Materiales y métodos

### Instrumento de medición

Se recurrió al cuestionario estructurado de tal forma que el resultado fuera un indicador que permitiera conocer el conocimiento que tiene el público en general de la mercadotecnia a través de medios electrónicos. Fue elaborado en Google Forms, lo que facilitó su dispersión, llegar a una mayor cantidad de clientes potenciales. En él se vieron reflejados los métodos cuantitativo y cualitativo.

### Método de muestreo y aplicación

Una vez definido el instrumento de recopilación (encuesta), su aplicación se realizó usando un muestreo no probabilístico. El muestreo no probabilístico selecciona la muestra con parámetros previamente establecidos, con el fin de obtener la muestra más representativa del mercado a investigar. Por lo que se delimitó a gente en la región de los Altos de Jalisco; el acceso a ella se efectuó en línea. Se generó un filtro para la selección de los encuestados; así, se obtuvo una visión específica de la muestra. Su aplicación en línea facilitó la recopilación y extracción de datos, y sin costo, ya que se usó una plataforma gratuita. Los resultados abren la puerta a implementar acciones más rápidas y seguras como parte del proyecto. De esta manera, se prueba el instrumento y se descubre todo aquello que es interesante afinar y mejorar en el instrumento.

Su gran desventaja está relacionada con el poco interés que despierta en el encuestado, lo cual disminuye la tasa de respuesta y obliga a incrementar el tiempo de aplicación de esta.

También se corre el riesgo de que una misma persona la conteste más de una vez, y arrojar datos erróneos.

## Esquema de distribución del cuestionario

Para distribuir la encuesta y alcanzar la muestra representativa, se eligieron las tres empresas mencionadas previamente por ser fabricantes de embutidos de la región. A cada una de ellas se acudió para solicitar su ayuda para colocar dicha encuesta en su página de Facebook empresarial. A través de esta red social se tuvo gran respuesta, dado que los clientes actuales aparecen como contactos en las páginas.

Asimismo, la encuesta se hizo llegar vía mensaje directo al cliente frecuente, a través de redes sociales personales y WhatsApp. Su amplia distribución proporcionó una gran cantidad de respuestas, las cuales sobrepasaron la meta inicial de 150 (50 por empresa), lo que permitió considerar válidos los resultados; al final, se tomaron 60 encuestas por empresa.

## Resultados

Las preguntas aplicadas a través de la plataforma de Google fueron proporcionadas primeramente al encargado de cada empresa con el propósito de obtener información previa a la que se arrojaría con el *marketing* digital, y conocer hasta dónde se tendrán que aplicar estas metodologías y tecnologías en las empresas de la región para proyectar su permanencia y crecimiento.

- ¿Su empresa cuenta con alguna incursión en el *marketing* digital?
- ¿Considera que su empresa está preparada para afrontar estas herramientas?
- ¿Tiene personas capacitadas dentro de la empresa para ejecutarlo?
- ¿En qué considera que ayudaría su implementación?
- ¿Qué alcance cree que tendría el *marketing* digital en su empresa?
- ¿Hasta dónde estaría dispuesto a incursionar en el *marketing* digital?

**Figura 1.** Presencia del *marketing* digital en la empresa

¿Su empresa cuentas con alguna incursión en el marketing digital?
1 respuesta

Por el momento no, ya que no cuenta con un departamento en esa área

¿Considera que su empresa esta preparada para afrontar estas herramientas?
1 respuesta

Si, pero con asesoría de alguien especializado

¿Tiene personas capacitadas dentro de la empresa para ejecutarlo?
1 respuesta

Por el momento no

¿En que considera que ayudaría la implementación del mismo?
1 respuesta

En poder crecer el negocio, y adquirir reconocimiento

¿Que alcance cree que tendría el Marketing Digital en su empresa?
1 respuesta

Tendría un alcance bueno, ya que ofrecemos un producto con una diferencia competitiva que puede interesar a más consumidores fuera de lo local.

¿Hasta donde estaría dispuesto a incursionar en el Marketing Digtial?
1 respuesta

Hasta donde la misma empresa lo permita, y el asesor nos ayude.

Fuente: Elaboración propia

En la figura 1 se aprecian los datos que nos entregaron como respuestas a las preguntas proporcionadas. Cabe precisar que no hubo certeza sobre quién fue el responsable del llenado. Aun así, permitió analizar el lugar dónde se encuentran actualmente y hacia dónde se dirigen.

## Opinión de los clientes

De la misma manera, se elaboró y aplicó el cuestionario dirigido al cliente de cada empresa participante; ello tuvo como objetivo conocer un poco cómo actúa y piensa el cliente sobre el canal usado en ese momento para que el producto les llegue, cómo se convirtió en cliente y su opinión sobre lo que pueda suceder al implementar el *marketing* digital en su relación cliente-empresa.

- ¿Cómo conoció los productos?
- ¿Sabes dónde más conseguir el producto, además del lugar en el que sueles comprarlo?
- ¿Alguna vez has buscado información relacionada con la empresa (punto de venta, ubicación, teléfono, dirección, etc.) a través de Internet?
- ¿Crees que te sería útil que la empresa realizara campañas de publicidad a través de Internet?

Los resultados se concentraron para su interpretación en una gráfica por pregunta.

**Figura 2.** Conocimiento del producto

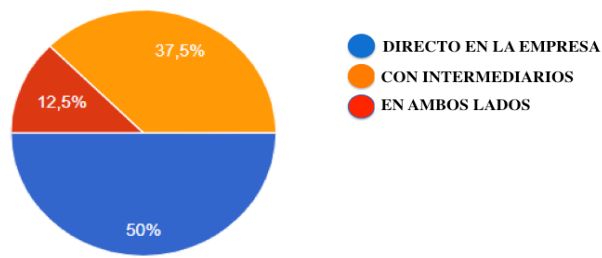

Fuente: Elaboración propia

Los resultados obtenidos en las tres empresas participantes se conjuntaron en una sola gráfica, ya que se trata de un ejercicio orientado a conocer el mayor o menor grado que existe en la región al incorporar las nuevas tecnologías mercadológicas en combinación con las tecnologías de comunicación.

Los resultados obtenidos reflejan que alrededor de 50 % de los encuestados conoció la empresa de forma directa por vivir en el área, por saber de la calidad del producto o a razón de que sus precios compiten con marcas nacionales y extranjeras; 37.5 % llegó a los productos a través de alguien que se los obsequió inicialmente, y al resto alguien se los recomendó. Así, teniendo en cuenta lo anterior, se puede afirmar que la publicidad de boca en boca se suele utilizar en la mayoría de pequeñas empresas regionales, lo cual se debe básicamente a que el cliente ha sido su principal promotor, lo que obliga a mantener una buena relación cliente-empresa, ya que gracias a esta logran mantener y atraer más clientes.

**Figura 3.** Presencia en el mercado

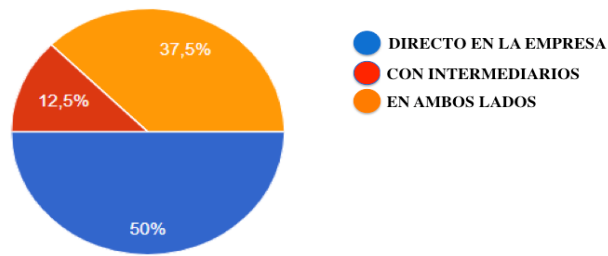

Fuente: Elaboración propia

Ahora bien, 50 % de las encuestas muestra que el cliente compra directamente en la empresa; 37.5 % de los clientes conocen otro lugar además de la empresa en donde los adquieren, y 12.5 % acude, por diversas circunstancias, con intermediarios (pequeñas tiendas de abarrotes que los venden).

El alto porcentaje de adquisición directa con las empresas por contar con distintos puntos de venta obviando intermediarios es algo en lo que han sido muy enfáticas las empresas, de esta manera el cliente evita desplazarse al lugar de producción para adquirirlos, además de que usan carteles para anunciarlos y la distribución de volantes que anuncian sus ubicaciones. Sus primeros ensayos en el uso de medios de *marketing* digital quedaron plasmados en las cuentas de Facebook abiertas para informar las promociones. Si bien estos esfuerzos no han sido bien comandados, con ellos han aumentado un poco sus clientes y ventas.

**Figura 4.** Puntos de venta

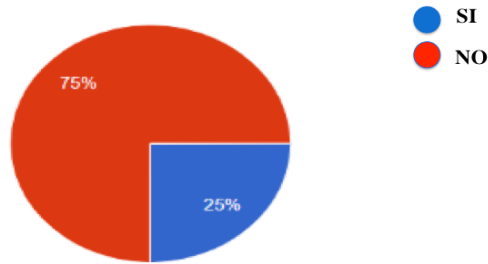

Fuente: Elaboración propia

**Figura 5.** Búsqueda a través de Internet

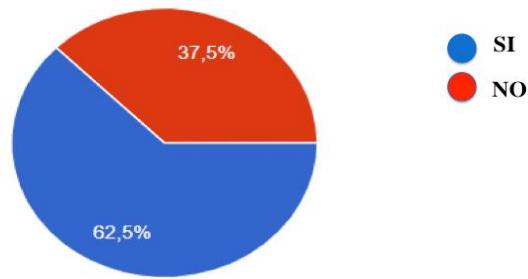

Fuente: Elaboración propia

Aunado a lo anterior, 62.5 % de los encuestados afirmaron que han buscado información relacionada con las empresas de embutidos aquí mencionadas. Con ello se muestra que hay un gran interés por parte de empresas y clientes por los servicios en medios electrónicos.

**Figura 6.** Publicidad en Internet

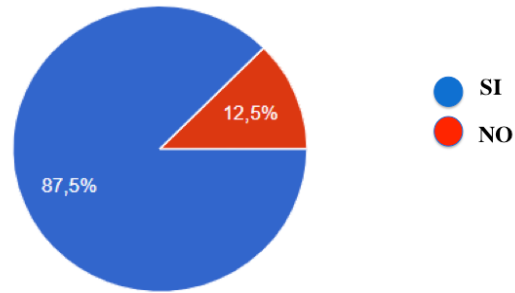

Fuente: Elaboración propia

En esa misma línea, 87.5 % de los encuestados considera importante que las empresas de embutidos realicen campañas de publicidad a través de Internet. Es común tener un teléfono inteligente conectado a Internet, y se puede buscar infinidad de productos y empresas. Si bien la publicidad de boca en boca sigue atrayendo clientes a las empresas, y son de edades distintas, lo cual ha motivado a continuar confiando en ella, es momento también de que las empresas cuenten con sus propias aplicaciones (*apps*), ya que el *marketing* digital incrementa el número potencial de clientes, no solo del área, pero con optimismo, basados en la calidad de los productos elaborados, en el futuro cercano, se podrá pensar en expansiones a poblaciones con mayor densidad poblacional y, por ende, clientes en potencia.

Con ello se darían a conocer los puntos de venta en distintas poblaciones y, posteriormente, pensar en entregas directas del fabricante al consumidor, e incluso poder generar una relación más cercana entre cliente-empresa.

## Discusión

Las entrevistas descritas arrojaron información valiosa para que las empresas consideren muy seriamente implementar el *marketing* digital, y con ayuda de esta herramienta ampliar sus mercados. Al adoptar estas técnicas mercadológicas modernas que abrirán el paso al desarrollo de la región, tendrán que emplear nuevo personal para efectuar estudios mercadológicos, sus empleados podrán asegurar sus trabajos por mayor tiempo, y en general, las empresas de la región, al emplear el *marketing* digital, lograrán un mejor posicionamiento de marca y productos.

Como herramienta mercadológica, el *marketing* digital agrupa las estrategias de mercadeo que usa la Web para que el usuario concrete su visita, todo ello basado en acciones planeadas previamente en cada empresa. Va mucho más allá de las formas tradicionales de venta y mercadeo conocidas; se trata de integrar estrategias y técnicas diversas pensadas exclusivamente para el mundo digital (Selman, 2017).

La publicidad de boca en boca es de alcance limitado. Al confiar exclusivamente en ella, el fabricante de embutidos desaprovecha las ventajas de los medios digitales y, por ende, restringe su crecimiento. Lo anterior se establece por el hecho de que en la actualidad una gran cantidad de gente interactúa frecuentemente con diferentes herramientas que ofrece el Internet, ya que con ellas es muy sencillo contactar con la gente en tiempo real, por lo que se ha convertido en un método más eficiente para dar seguimiento al negocio desde equipos portátiles; en suma, la audiencia es mayor a un costo mínimo.

Es un hecho, según Real, Leyva y Heredia (2014), que 80 % de las empresas que tienen un aplazamiento en la tecnología, y no están a la vanguardia de las nuevas estrategias de *marketing* y ventas, desaparecen con el tiempo.

Para garantizar la presencia en el mercado se requiere adoptar la innovación. Hoy en día, las empresas tienen que mejorar e implementar estrategias digitales, con el fin de estar en ventaja competitiva por encima de sus contendientes, lograr un crecimiento sostenible y asegurar su posicionamiento en el mercado.

## A manera de conclusión

Las evidencias muestran que las pequeñas empresas alteñas siguen utilizando métodos de publicidad tradicionales como el de boca en boca. Después de entrevistar a diversos grupos de la región, a saber, clientes de las empresas fabricantes de embutidos, docentes, personas especializadas en *marketing* digital y otros ciudadanos de los Altos de Jalisco, se concluye que muchas de las empresas de la región nunca han recurrido al *marketing* digital, por lo que están desaprovechando las herramientas que han estado surgiendo en los últimos años. Según los resultados aquí obtenidos, 85.5 % de los encuestados cree que las empresas deben acceder a los medios digitales para proporcionar publicidad e información que facilite el contacto directo con la empresa y sus productos, ahorrar tiempo, costos, tener un mayor alcance, lograr un mejor posicionamiento en su mercado y, con todo ello, buscar estar mejor que sus competidores.

Alrededor de 62.5 % de los entrevistados dijo haber buscado alguna vez información relacionada con las empresas a través de los medios digitales. Además, indicaron que actualmente la gente pasa cada vez mayor tiempo conectada a las redes sociales, y aquella empresa que no usa esas herramientas se pone en desventaja, ya que todo cliente espera encontrar todo lo relacionado con sus gustos y preferencias en el mínimo tiempo.

La permanencia de la empresa en un mundo tan versátil implica innovación y adaptación al uso de nuevos procesos administrativos. Los requerimientos de la demanda actual y la creación de nuevos competidores reclama a las empresas con historia usar las nuevas tecnologías para beneficio propio y de sus clientes; una resistencia a incorporar las nuevas tendencias impedirá el desarrollo y crecimiento de la organización.

Cabe mencionar que, aunado a lo aquí ya mencionada, también se estará desaprovechando el potencial que tienen los egresados de distintas universidades, a donde acuden a formarse jóvenes para orientar a las empresas y hacerlas crecer. Sin duda, aquí está en juego el crecimiento de la región como nunca antes. Asimismo, con estás nuevas metodologías se dejará de contribuir al ingreso estatal como productor de materias primas, tal como lo es en la actualidad, para convertirse en fabricantes de productos elaborados de gran calidad, así lo son las materias primas que se emplean en estos productos y tantas otras que se exportan a otras regiones del país.

Todo ello dará cabida a una gran cantidad de profesionistas de otras áreas, como los profesionales de las tecnologías de la información, ya que son base para desarrollar las mencionadas aplicaciones que permitan conocer los productos en cualquier parte del mundo. Y así, paulatinamente, se irán abriendo las puertas para que las empresas crezcan sin perder sus métodos tradicionales de producción, ya que de estos depende el sabor de sus productos. Por último, también habrá que incorporar gente del área creativa para presentar de maneras distintas las ideas, así como gente del área de negocios internacionales para que puedan ir visualizando los mercados que se podrán abarcar.

## Referencias